\documentclass[11pt,a4paper]{article}

\usepackage[margin=2.5cm]{geometry}
\usepackage{color}
\usepackage{amsmath,amssymb,amsthm,amsopn,bm,mathrsfs}
\usepackage{setspace}
\usepackage{graphbox}
\usepackage[en-GB]{datetime2}
\DTMlangsetup[en-GB]{ord=omit}
\usepackage{natbib}
\usepackage[hidelinks]{hyperref} 

\DeclareMathOperator{\E}{\boldsymbol{\mathbb{E}}}
 
\DeclareMathOperator{\Prob}{\boldsymbol{\mathbb{P}}}

\newcommand{\argmin}[1]{\underset{#1}{\mathrm{argmin}} \ }
\newcommand{\argmax}[1]{\underset{#1}{\mathrm{argmax}} \ }
\newcommand{\D}{\mathsf{D}}

\newcommand{\bH}{{\bf H}}
\newcommand{\bbm}{{\bm m}}

\newcommand{\bx}{{\bm x}}

\newcommand{\bX}{{\bf X}}

\newcommand{\diff}{\mathsf{d}}
\renewcommand{\vec}{\operatorname{vec}}

\newcommand{\bbeta}{{\bm \beta}}
\newcommand{\pkg}[1]{{\tt #1}}
\newcommand{\proglang}[1]{{\sf #1}}
\newcommand{\code}[1]{{\tt #1}}

\title{Statistical visualisation of tidy and geospatial data in R via kernel smoothing methods in the eks package}

\author{Tarn Duong\footnote{Paris, France F-75000. Email:
{\tt tarn.duong@gmail.com}}}

\begin{document}

\maketitle

\begin{abstract}
\noindent Kernel smoothers are essential tools for data analysis due to their ability to convey complex statistical information with concise graphical visualisations. Their inclusion in the base distribution and in the many user-contributed add-on packages of the \proglang{R} statistical analysis environment caters well to many practitioners. Though there remain some important gaps for specialised data, most notably for tidy and geospatial data. The proposed \pkg{eks} package fills in these gaps. In addition to kernel density estimation, this package also caters for more complex data analysis situations, such as density derivative estimation, density-based classification (supervised learning) and mean shift clustering (unsupervised learning). We illustrate with experimental data how to obtain and to interpret the statistical visualisations for these kernel smoothing methods.

\end{abstract}

\section{Introduction}

Kernel smoothers form an essential suite of statistical techniques for data analysis in the 21st century due to their ability to convey complex statistical information in a concise and intuitive visual format. This ability arises from their shared characteristic of transforming data samples into smoothed estimates. Kernel smoothers have provided insight in data analysis problems in many situations. A small recent selection of these includes: the identification of important biomedical functions, such as characterising different sub-cellular structures in single cells \citep{schauer2010} or characterising a single cell population in mixed cell samples \citep{chacon2011ss}; the evaluation of predicted extreme temperatures to calibrate  climate models \citep{beranger2019}; the estimation of the home range of animal movements \citep{baillo2021}; or the detection of traffic anomalies from traffic flows \citep{kalair2021}. 

A major access point to kernel smoothers in the \proglang{R} statistical programming environment is the \pkg{ks} (`\pkg{k}ernel \pkg{s}moothing') add-on package \citep{duong2007}, which implements density estimation, density derivative estimation, classification (unsupervised learning), clustering (unsupervised learning), and inferential methods. This package utilises the base \proglang{R} graphics engine to generate its statistical graphics. Whilst it remains the most comprehensive graphics engine in \proglang{R}, the \pkg{ggplot2} graphics engine \citep{ggplot2} has gained popularity, as part of the `tidyverse', especially with data analysis practitioners.  Despite the dramatic rise in the number of analysis methods available in the tidyverse, nonetheless it comprises a limited range of natively implemented kernel smoothers. The first goal of the \pkg{eks} (`\pkg{e}xtended \pkg{k}ernel \pkg{s}moothing') package \citep{eks} is to provide access to a comprehensive suite of kernel smoothers in the tidyverse.
 
There is an analogous lack of kernel smoothers for geospatial data analysis. Since the term `geospatial' data analysis refers to many different yet overlapping concepts, we employ it in this paper to refer to data analysis which is compatible with `Geographical Information Systems' (GIS). Within \proglang{R}, the \pkg{sf} package \citep{sf} provides geospatial/GIS functionality via its robust implementation of the `simple features' GIS standard data format \citep{ogc-sfa}. The  \pkg{eks} package relies on this simple features implementation, in order to facilitate visualisations in both \pkg{ggplot2} and base \proglang{R} graphical engines, and input/output to external GIS software (such as ArcGIS and QGIS).  The second goal of the \pkg{eks} package is to provide access to a comprehensive suite of kernel smoothers for geospatial analysis. 

\begin{figure}[!ht]
\centering
\begin{tabular}{@{}c@{}c@{}}
\includegraphics[width=0.465\textwidth,align=t]{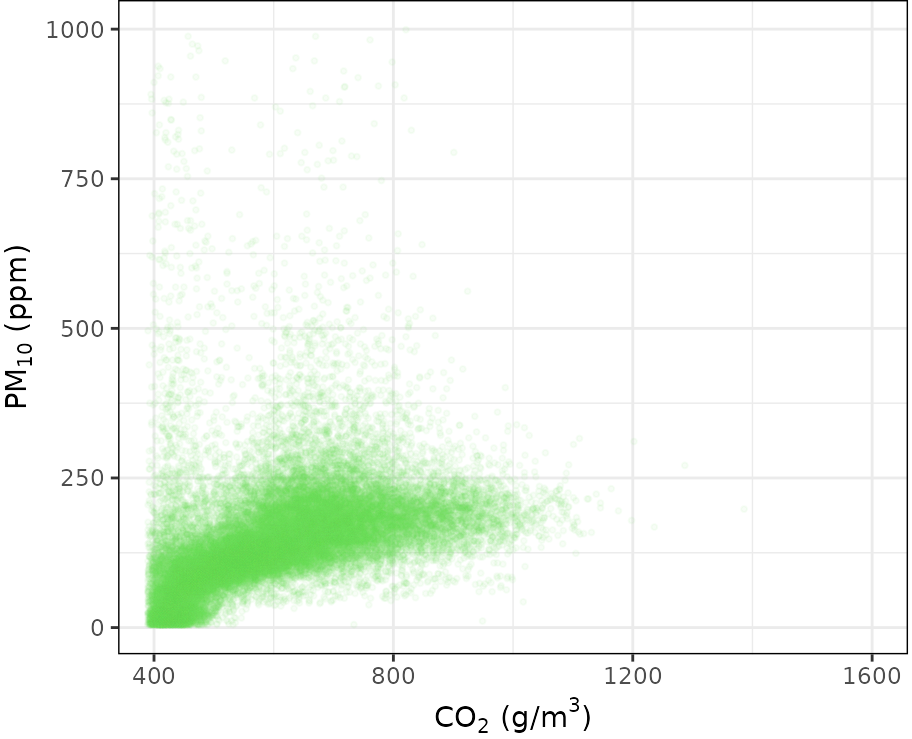} & \, 
\includegraphics[width=0.4\textwidth,align=t]{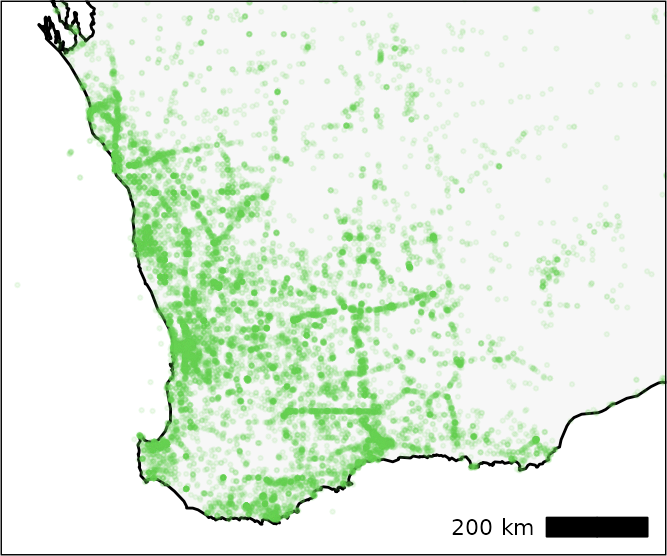} 
\end{tabular}
\caption{Scatter plots of data sets. (Left) Tidy air quality measurements \code{air} ($n=30239$), in the Ch\^atelet underground train station in Paris, France.  (Right) Geospatial {\it Grevillea} locations \code{grevilleasf} ($n=22203$), in the biodiversity hotspot of south-western Western Australia.}
\label{fig:scatter}
\end{figure}

Thus a wide range of kernel smoothers is now available for tidy and geospatial data, and for \pkg{ggplot2} and base \proglang{R} graphical visualisations. The user is able to select and combine these components, with their differing strengths and applicabilities, in order to construct suitable data analysis workflows. To illustrate kernel smoothers, we employ the tidy data set \code{air} from the \pkg{ks} package, and the geospatial data set \code{grevilleasf} from the \pkg{eks} package, as shown in Figure~\ref{fig:scatter}. 

A tidy data set is a data matrix where (i) each variable forms a column and (ii) each observation forms a row, and it is also known as a `long' data set \citep{wickham2014}. The first three records of the \code{air} data set are
\begin{verbatim}
R> data(air, package="ks")
R> air
        date  time no no2 pm10 co2 temp humi
1 2013-01-01 01:00  6  31  182 776 21.5 46.2
2 2013-01-01 02:00  6  30  166 800 21.6 47.3
3 2013-01-01 03:00  4  27  124 799 21.8 47.0
...
\end{verbatim} 
These are the hourly mean air quality measurements from 01 January 2013 to 31 December 2016 in the Châtelet underground train station, which is a major hub in the Paris transport network   \citep{ratp2016}. We focus on the concentrations of carbon dioxide CO$_2$ (g/m$^3$) (\code{co2}) and of particulate matter less than 10 $\mu$m in diameter PM$_{10}$ (parts per million) (\code{pm10}), and the hourly interval (\code{time}). The concentrations of CO$_2$ indicate the renewal rate of fresh air, and of PM$_{10}$ the potential to affect adversely respiratory health. There are $n=30239$ complete (\code{co2}, \code{pm10}) measurements. 

The geospatial data \code{grevilleasf} data set consists of 22203 plants from 238 different {\it Grevillea} species collected in Western Australia. The south-west corner of Western Australia is one of the 25 `biodiversity hotspots' which are `areas featuring exceptional concentrations of endemic species and experiencing exceptional loss of habitat' identified in \citet{myers2000} to assist in formulating priorities in biodiversity conservation policies. The geodetic coordinates (degrees) of the {\it Grevillea} locations are transformed into planar coordinates (metres) using the GDA2020/MGA zone 50 (EPSG:7850) projection. They are encoded as a simple feature in the column \code{geometry}, which is a special data structure that cannot be treated like the usual floating point variables in data frames or tibbles, and require specialised methods implemented by the \pkg{sf} package. So we refer to \code{grevilleasf} solely as a geospatial data set, and omit any mention of its tidy status, to emphasise its distinct geospatial characteristics.    
\begin{verbatim}
R> data(grevilleasf, package="eks")
R> grevilleasf
Simple feature collection with 22303 features and 2 fields
Geometry type: POINT
Dimension:     XY
Bounding box:  xmin: 73519.97 ymin: 6120859 xmax: 1795868 ymax: 8451928
Projected CRS: GDA2020 / MGA zone 50
                  name   species                 geometry
1    Grevillea robusta   robusta POINT (390106.5 6462671)																																																									
2   Grevillea speciosa  speciosa POINT (382689.2 6457387)
3    Grevillea robusta   robusta POINT (390089.8 6462603)
...
\end{verbatim}

This paper focuses on the software implementation of the kernel smoothers, and is complementary to \citet{chacon2018} which focuses on the underlying statistical framework. The \pkg{eks} package computes kernel smoothers for 1- and 2-dimensional tidy data, and 2-dimensional geospatial data. In Section~\ref{sec:kde}, we explore kernel density estimation, in Section~\ref{sec:kda} classification (supervised learning), in Section~\ref{sec:kdde} density gradient estimation, and in Section~\ref{sec:kms} clustering (unsupervised learning). We illustrate each case first for tidy data with \code{ggplot2} graphics, and then for geospatial data with \code{ggplot2} and base \proglang{R} graphics. In Section~\ref{sec:other}, we  briefly mention kernel smoothers in other data analysis settings, which are implemented in the \pkg{eks} package but have been omitted for brevity, and we end with some concluding remarks.

\section{Density estimation} \label{sec:kde}

Density estimation is a fundamental statistical analysis tool, since it supplies much information about the data set at hand. Our data $\bX_1, \dots, \bX_n$ is a random sample drawn from the common density function $f$. The goal of density estimation, as its name suggests, is to estimate this unknown density. Kernel density estimates are a popular choice among the many available smoothed density estimation methods, since they possess an intuitive construction. It is the most widely used kernel smoother, and can be considered to be a smoothed version of the histogram. For an arbitrary estimation point $\bx$, the kernel density estimate is  
\begin{equation}
\hat{f}_\bH(\bx) = n^{-1} \sum_{i=1}^n K_\bH(\bx - \bX_i).
\label{eq:kde}
\end{equation} 
Throughout the \pkg{eks} package, the kernel function is the Gaussian density function $K_\bH (\bx) = (2\pi)^{-1} |\bH|^{-1/2} \exp (-\tfrac12 \bx^\top \bH^{-1} \bx)$.  Equation~\eqref{eq:kde} tells us that to compute a kernel density estimate, we place a Gaussian function, with variance $\bH$, at each data point $\bX_i$, and then we sum these kernel functions. This way, the data sample $\bX_1, \dots, \bX_n$ are transformed into a smooth surface $\hat{f}_\bH$. \citet[Chapter~2]{chacon2018} contains a more detailed overview of kernel density estimates. 

The bandwidth matrix $\bH$ in Equation~\eqref{eq:kde} is the crucial tuning parameter. A bandwidth matrix which is too small leads to an undersmoothed density estimate since it does not offer sufficient reduction in the complexity of the observed data. On the other hand, a bandwidth matrix which is too large leads to an oversmoothed density estimate that obscures important details in the observed data. Thus it is critical to find an optimal trade-off between this under- and oversmoothing. Many possible solutions for optimal smoothing are implemented in the \pkg{ks} package, and are thus available in the \pkg{eks} package, including the plug-in, unbiased cross validation and smoothed cross validation bandwidths. These selectors are implemented solely for the Gaussian kernel since it ``allows for important mathematical and computational simplifications and avoids any possible problems with the non-existence of higher order derivatives of the kernel function when computing data-based bandwidth selectors'' \citep[p.~82]{chacon2018}.

\subsection{Tidy density estimation}

\begin{figure}[!ht]
\centering
\includegraphics[width=\textwidth]{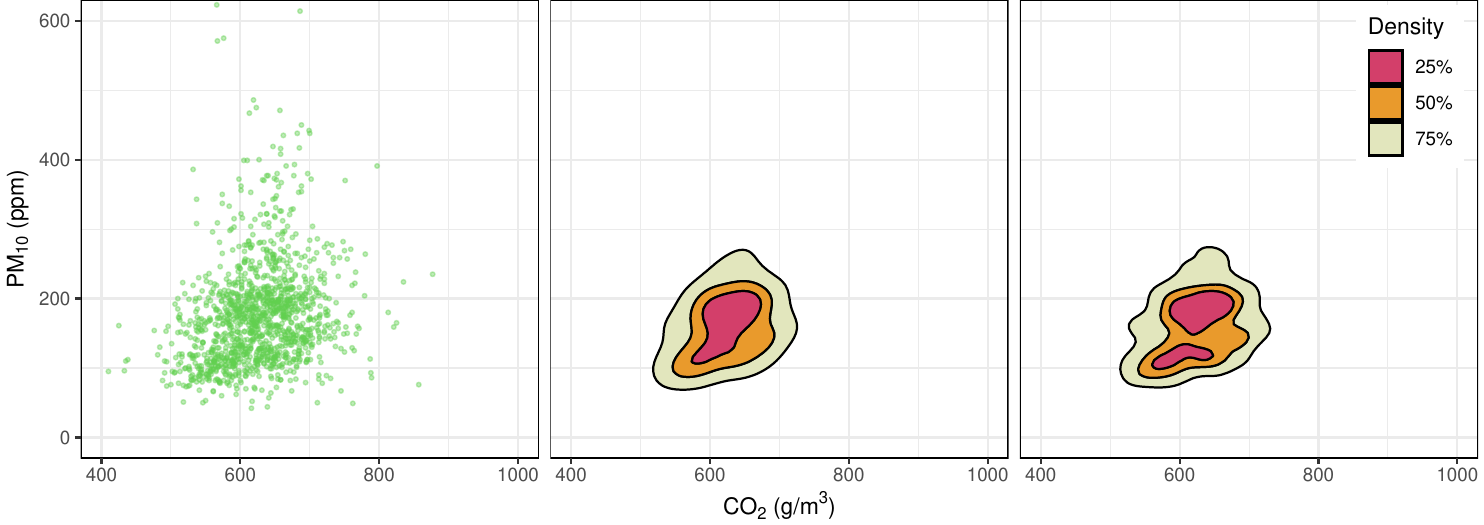} 
\caption{Filled contour plots of density estimates for air quality measurements 11:00--12:00 ($n=1285$) with quartile probability contour levels. (Left) Scatter plot. (Centre) Optimally smoothed. (Right) Undersmoothed.}
\label{fig:kde1}
\end{figure}

To illustrate density estimation for tidy data, we focus on a single hourly interval of the air quality measurements. Figure~\ref{fig:kde1} compares the density estimates, for the $n=1285$ measurements from 11:00 to 12:00, with an optimal bandwidth and a sub-optimal one. The optimal bandwidth is computed from the \pkg{eks} package and the sub-optimal one from the \pkg{ggalt} package \citep{ggalt}. The former, known as the bivariate plug-in bandwidth matrix \citep{duong2003}, is the default optimal bandwidth in the \pkg{eks} package, and it is obtained from a call to \code{ks::Hpi}. This optimality is the result from theoretical and numerical comparisons in \citet[Section~2.3]{chacon2018} and the references therein. For the air quality measurements, the optimal \code{ks::Hpi} matrix is [342.1, 97.2; 97.2 365.2]. The presence of non-zero off-diagonal entries in the optimal matrix appropriately orients the kernel functions, and the resulting density estimate is unimodal, as shown in the centre panel of Figure~\ref{fig:kde1}. The default bandwidth in \pkg{ggalt}, which is widely used in the tidyverse, is obtained from the element-wise application of the univariate plug-in bandwidth \code{KernSmooth::dpik}. For the air quality measurements, this bandwidth is [218.6, 0; 0, 124.9]. Since this sub-optimal matrix only applies smoothing in the coordinate axis directions, it yields an undersmoothed density estimate with spurious bimodal structure on the right panel. 

In Figure~\ref{fig:kde1}, the heights of the contour regions are calculated according to the probability contours method \citep{bowman1993,hyndman1996jasa}. The pink region is the smallest region that contains 25\% of the probability mass, the orange region plus the enclosed pink region is the smallest region that contains 50\% of the probability mass, and the yellow region plus the enclosed orange and pink regions is the smallest region that contains 75\% of the probability mass. Since these are relative heights, they facilitate the choice of the contour levels, since it involves selecting values from 0\% to 100\%, rather than from the range of the density values. These probability contours can also be considered as a multivariate extension of the univariate percentiles, e.g., the 50\% contour region is a bivariate equivalent to the median. Due to their intuitive properties, these probability contours are employed throughout in \pkg{eks}, with the quartile contour levels (25\%, 50\%, 75\%) being the default values. In addition to their intuitive interpretation, these probability contours are straightforward to compute: the kernel density estimate is evaluated at the $n$ observed data values $\hat{f}_\bH(\bX_1), \dots, \hat{f}_\bH(\bX_n)$, then we compute $\tau_\alpha$ as the $\alpha$-quantile of these evaluated values, and the $\alpha$ probability contour region is the level set of the density estimate at $\tau_\alpha$, i.e. $\{\bx: \hat{f}_\bH(\bx) > \tau_\alpha\}$ \citep{hyndman1996jasa}. These probability contours are also implemented in the \pkg{ggdensity} package \citep{ggdensity}, though with a similar sub-optimal bandwidths as in \pkg{ggalt}.  

The \proglang{R} code snippets included here are intended to give an overall idea of the syntax of the \pkg{eks} package, rather than a complete code to reproduce the figures. The latter is provided in the companion \proglang{R} script. The code snippet to compute the density estimate with the optimal bandwidth, in the centre panel in Figure~\ref{fig:kde1}, is
\begin{verbatim}
R> ## tidy density estimate
R> air2 <- ungroup(filter(air, time=="11:00"))
R> air2 <- select(air2, co2, pm10)
R> t1 <- tidy_kde(air2, H=H1)
R> ggplot(t1, aes(x=co2, y=pm10)) + geom_contour_filled_ks(colour=1)
\end{verbatim}
The function \code{tidy\_kde} is a wrapper function for \code{ks::kde}, which computes the tidy density estimate explicitly. This differs from existing layer functions, e.g., \code{ggplot2::geom\_density\_2d} and \code{ggalt::geom\_bkde2d}, which compute the density estimate internally and do not return a user-level \proglang{R} object. The tidy density estimate from \code{tidy\_kde} is:
\begin{verbatim}
R> t1
# A tibble: 22,801 x 6
     co2  pm10  estimate ks        tks   label  
   <dbl> <dbl>     <dbl> <list>    <chr> <chr>  
 1  342. -28.8 -5.65e-24 <kde>     kde   Density
 2  346. -28.8 -7.42e-22 <int [1]> kde   Density
 3  350. -28.8  2.38e-22 <int [1]> kde   Density
...
\end{verbatim}
This output is a tidy tibble with an added \code{tidy\_ks} class, which allows for a \code{ggplot.tidy\_ks} method to be defined for this object class. Otherwise, it can be treated as a tibble. The first two columns \code{co2, pm10} (same names as the input data) are the coordinates of the vertices in the estimation grid, the third column \code{estimate} is the density estimate value at \code{co2, pm10}. The fourth column \code{ks} holds the output from \code{ks::kde}. This is required for the computation of probability contours in the new layer function \code{geom\_contour\_filled\_ks} to draw the filled contour plots for \code{tidy\_ks} objects. The remaining columns indicate that the output is a density estimate computed from \code{ks::kde}, and they are employed in \code{ggplot.tidy\_ks} to create default aesthetic mapping and legend labels. This default aesthetic mapping is \code{ggplot2::aes(x=co2, y=pm10, z=estimate, weight=ks)}. Whilst the \code{x}, \code{y}, \code{z} aesthetics are as expected for a bivariate contour plot, the \code{weight} aesthetic is unorthodox, since it is not a weighting variable: it is a workaround in \pkg{ggplot2} graphics to mimic the dynamic display of probability contours in base \proglang{R} graphics.     

For the air quality measurements from 11:00 to 12:00, the quartile contour levels for the optimally smoothed density estimate in the centre panel in Figure~\ref{fig:kde1} are 3.31e-5, 2.42e-5,  1.22e-5, and for the undersmoothed density estimate in the right panel are 3.51e-5, 1.22e-5, 2.53e-5. These probability contour heights are different for each different density estimate, even if the target contour probabilities remain the same at 25\%, 50\%, 75\%. On the other hand, it is sometimes useful to have a set of fixed contour heights for all density estimates for a direct comparison. A heuristic method consists of computing the probability contour heights for each density estimate, for a fixed set of probabilities, which are then aggregated. We compute the corresponding probabilities for each density estimate for this aggregated set of contour heights, and we remove any contour levels whose estimated probability which are too close to each other. This procedure is implemented in the \code{contour\_breaks} function, though some trial and error is  still likely required to produce visually appealing contour plots for all density estimates \citep[Section~2.2]{chacon2018}.  

\begin{figure}[!ht]
\centering
\includegraphics[width=0.67\textwidth]{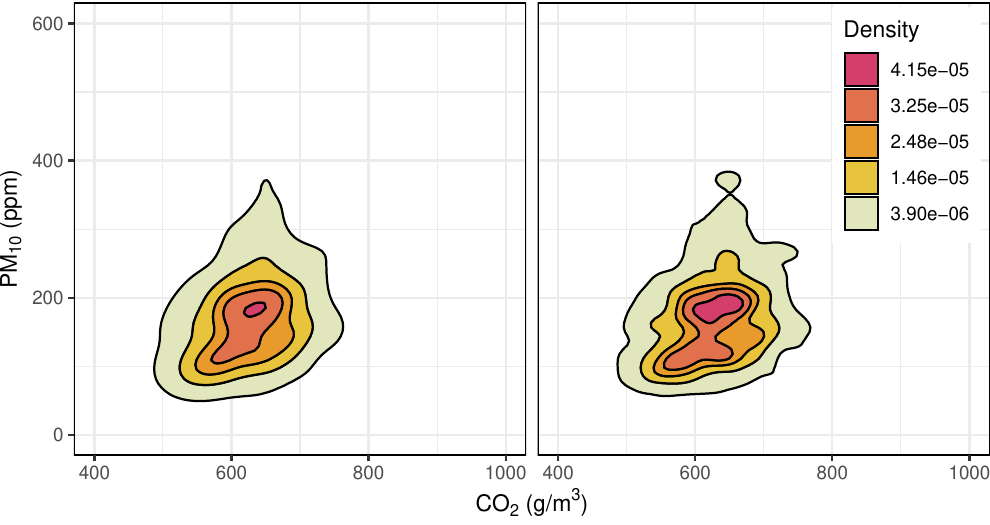} 
\caption{Filled contour plots of density estimates for air quality measurements from 11:00 to 12:00 ($n=1285$) with fixed contour levels. (Left) Optimally smoothed. (Right) Undersmoothed.}
\label{fig:kde2}
\end{figure}

We revisit the density estimates for the air quality measurements from 11:00 to 12:00, this time with the fixed contour heights (3.90e-6, 1.46e-5, 2.48e-5, 3.25e-5, 4.15e-5) in Figure~\ref{fig:kde2}. With these fixed contour heights, a direct comparison of different density estimates is possible. The optimally smoothed estimate on the right exceeds the two highest contour heights (4.15e-5 dark pink, 3.25e-5 dark orange) with a unimodal bump, whereas the density estimate on right exceeds them with a bimodal structure. Since the latter adds some spurious modal information, it is considered to be undersmoothed in comparison. 

The code to produce two density estimates with a single set of contour heights in  Figure~\ref{fig:kde2} is 
\begin{verbatim}
R> ## fixed contour levels
R> H2 <- diag(sapply(air2, KernSmooth::dpik)^2)
R> t2 <- tidy_kde(air2, H=H2)
R> t3 <- c(t1, t2)
R> b <- contour_breaks(t3, cont=c(10,30,50,70,90))
R> ggplot(t3, aes(x=co2, y=pm10)) + geom_contour_filled_ks(colour=1, breaks=b) 
+  facet_wrap(~group)
\end{verbatim}

\subsection{Geospatial density estimation}

\begin{figure}[!ht]
\centering
\includegraphics[width=0.95\textwidth]{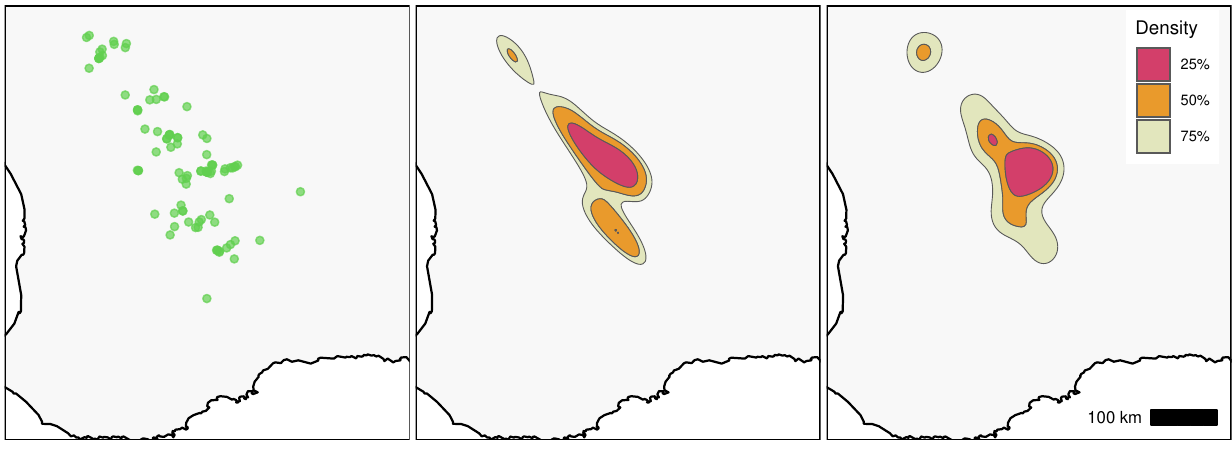} 
\caption{Filled contour plots of density estimates for {\it G. yorkrakinensis} ($n=93$) with quartile probability contour levels. (Left) Scatter plot. (Centre) Optimally smoothed. (Right) Oversmoothed.}
\label{fig:kde3}
\end{figure}

To illustrate density estimation for geospatial data, we utilise single species subsets of the {\it Grevillea} locations. Figure~\ref{fig:kde3} compares the density estimates for the $n=93$ locations of the {\it G. yorkrakinensis} species which result from an optimal plug-in bandwidth [8.84e8, $-8.33$e8; $-8.33$e8, 1.36e9] and a sub-optimal bandwidth [5.43e8, 0; 0, 9.10e8]. The optimally smoothed density estimate in the centre panel in Figure~\ref{fig:kde3} displays a trimodal structure with obliquely oriented contours. The oversmoothed density estimate in the right panel, whilst it also has trimodal structure, it has circular contours which do not follow closely the orientation of the observed data points, and so is considered to be oversmoothed. 

To produce the centre panel in Figure~\ref{fig:kde3} for the {\it G. yorkrakinensis} locations, the commands are 
\begin{verbatim}
R> ## geospatial density estimate 
R> yorkr <- filter(grevilleasf, species=="yorkrakinensis")
R> s1 <- st_kde(yorkr)
\end{verbatim}
The function \code{st\_kde} is the geospatial equivalent of \code{tidy\_kde}, and produces an object of class \code{sf\_ks}, which is a list of 3 fields: \code{tidy\_ks}, \code{grid}, and \code{sf}. The first field is a summary of the tidy density estimate from \code{tidy\_kde}, the second are the rectangular polygons of the estimation grid, and the third are the 1\% to 99\% probability contour regions of the density estimate. We focus on the contour regions.
\begin{verbatim}
R> s1$sf
Simple feature collection with 99 features and 2 fields
Geometry type: GEOMETRY
Dimension:     XY
Bounding box:  xmin: 429181.6 ymin: 6333015 xmax: 757134 ymax: 6793115
Projected CRS: GDA2020 / MGA zone 50
   contlabel     estimate                       geometry
1         99 2.491961e-12 POLYGON ((430965 6755235, 4...
2         98 3.699348e-12 POLYGON ((437905.1 6746871,...
3         97 6.714617e-12 POLYGON ((448315.2 6740434,...
...
\end{verbatim}
This has 2 attributes: \code{contlabel} (label of probability contour region) and \code{estimate} (height of probability contour region). Unlike for \code{tidy\_kde} where the probability contour regions are computed dynamically in the layer function \code{geom\_contour\_filled\_ks}, these 1\% to 99\% regions are converted to (multi)polygons prior to plotting since the dynamic conversion during plotting could be computationally heavy. Since we are unable replicate the automatic selection of the quartile contours 25\%, 50\%, 75\% by default, like in \code{geom\_contour\_filled\_ks}, for the \code{ggplot2::geom\_sf} layer function, we first apply \code{st\_get\_contour} to the input of \code{ggplot2::geom\_sf}. The \code{sf\_ks} class also has a \code{ggplot.sf\_ks} method which computes the default map legend. 
\begin{verbatim}
R> ## geospatial density estimate geom_sf plot
R> ggplot(s1) + geom_sf(data=st_get_contour(s1), aes(fill=contlabel)) 
\end{verbatim}.
The following command produces the equivalent in base \proglang{R} graphics to the \code{ggplot2} plot in the centre panel in Figure~\ref{fig:kde2}.  
\begin{verbatim}
R> ## geospatial density estimate base R plot
R> plot(s1)
\end{verbatim}
This \code{plot.sf\_ks} method for \code{sf\_ks} objects method internally calls \code{st\_get\_contour} to extract the required contour polygons for plotting, so it is more concise than \code{ggplot2::geom\_sf} that requires an explicit user-level call to \code{st\_get\_contour}. 
The base \proglang{R} and \pkg{ggplot2} plots are essentially identical since they comply with the geospatial standard specifications for simple features.

\subsection{Optimal bandwidth matrices}

Since the bandwidth matrix is the crucial tuning parameter for kernel density estimates, we explore further their statistical properties. These properties are the subject of a vast body of research literature, which we do not attempt to review here, and instead provide a simplified outline of how the optimal bandwidth matrix in \pkg{eks} is obtained. 

We begin with a squared error discrepancy between a density estimate $\hat{f}_\bH$ and the target density $f$, i.e.,  $M(\bH) = \int \E [\hat{f}_\bH (\bx) - f(\bx)]^2 \, \diff \bx$. Since this expression involves the unknown target density $f$, it must be estimated for it to be of practical use. The plug-in bandwidth matrix in \code{ks::Hpi} computes the estimate $\hat{M}(\bH) = (4\pi)^{-d/2} n^{-1} |\bH|^{-1/2} + \tfrac14 \hat{\bbm}_4^\top (\vec \bH \otimes \vec \bH)$.  We omit to describe this estimate rigorously since it would require lengthy technical definitions: the interested reader is encouraged to consult \citet[Chapter~3]{chacon2018} for details. We are content to state that the first term in $\hat{M}$ is related to the variance of the density estimate, and the second term to the square of the bias of the density estimate. An optimal bandwidth matrix $\hat{\bH}$ is defined as 
\begin{equation}
\hat{\bH} = \argmin{\bH} \hat{M} (\bH)
\label{eq:H}
\end{equation}  
where the minimisation is carried out over the space of all symmetric positive definite matrices. When this minimisation is achieved, then there is an optimal trade-off between the variance and the squared bias, or equivalently between over- and under-smoothing. When an optimal bandwidth matrix $\hat{\bH}$ is substituted into Equation~\eqref{eq:kde}, the resulting kernel density estimate is the closest to the target density $f$ as measured by the discrepancy $\hat{M}$. 
Different bandwidth matrices arise from the different ways of computing $\hat{M}$ and/or from different ways of carrying out the minimisation. For example, the default bandwidth in \pkg{ggalt} treats the joint bivariate optimisation in Equation~\eqref{eq:H} as two separate univariate optimisation problems. The density estimate functions \code{tidy\_kde} and \code{st\_kde} compute $\hat{\bH}$ in Equation~\eqref{eq:H} by calling the \code{ks::Hpi} function, and then substitute this $\hat{\bH}$ into Equation~\eqref{eq:kde}, to compute an optimal tidy/geospatial density estimate, as shown in the centre panels in Figures~\ref{fig:kde1} and \ref{fig:kde2}. 

Additional bandwidth matrices in the \pkg{ks} package include the normal scale \code{ks::Hns}, unbiased cross validation \code{ks::Hucv} and smoothed cross validation \code{ks::Hscv}. The commands are:
\begin{verbatim}
R> ## smoothed cross validation selector
R> H3 <- ks::Hscv(air2)
R> t3 <- tidy_kde(air, H=H3)
\end{verbatim}
For most data samples, the plug-in bandwidth \code{ks::Hpi} yields fast and robust kernel estimates, though there remain some cases where other bandwidths are more suitable. For a review of the performance of these bandwidths, see \citet[Chapter~3]{chacon2018} and the references therein. For brevity, we illustrate kernel estimates only with the plug-in optimal bandwidths in the sequel.

\section{Density-based classification (supervised learning)} \label{sec:kda}

The goal of classification is to assign future data to one of the known classes in the current data. So this is a supervised learning problem. The data are $(\bX_1, Y_1), \dots, (\bX_n, Y_n)$, where the $\bX_i$ are the observed attributes, and the $Y_i$ is the known class label from  $m$ classes. These data are a random sample from the mixture density $\pi_1 f_1 + \dots + \pi_m f_m$, where $\pi_j$ is the prior probability and $f_j$ is the marginal density function for class $j$, for $j=1, \dots, m$ \citep[Section~7.2]{chacon2018}.  

The Bayes classifier assigns a candidate point $\bx$ to the class $c$ with the highest density value at $\bx$, i.e., $c(\bx) = \mathrm{argmax}_{j=1, \dots, m} \, \pi_j f_j(\bx)$. This Bayes classifier has few assumptions on the form of the target densities $f_j$ and achieves the smallest misclassification rate (Bayes error) among all classifiers given the attributes  \citep[p.~2]{devroye1996}. The misclassification error is the probability that we do not classify a candidate point in class $j$ given that it is drawn from class $j$, $\Prob \{c(\bX) \neq j | \bX \sim f_j\}$. 
The density-based classifier replaces the prior probability $\pi_j$ with the observed sample class proportion $\hat{\pi}_j$, and the marginal density $f_j$ with the marginal density estimate $\hat{f}_j$. Each marginal density estimate is computed with its own optimal bandwidth matrix. The estimated class label for $\bx$ from the kernel density-based classifier is thus $$\hat{c}(\bx) = \argmax{j=1, \dots, m} \hat{\pi}_j \hat{f}_j(\bx).$$
This kernel classifier is more adaptable than the usual linear and quadratic classifiers. The linear classifier uses Gaussian density fits with a common variance matrix for all classes, and the quadratic classifier Gaussian density fits with a different variance matrix for each class.

\subsection{Tidy classification}

\begin{figure}[!ht]
\centering
\includegraphics[width=\textwidth]{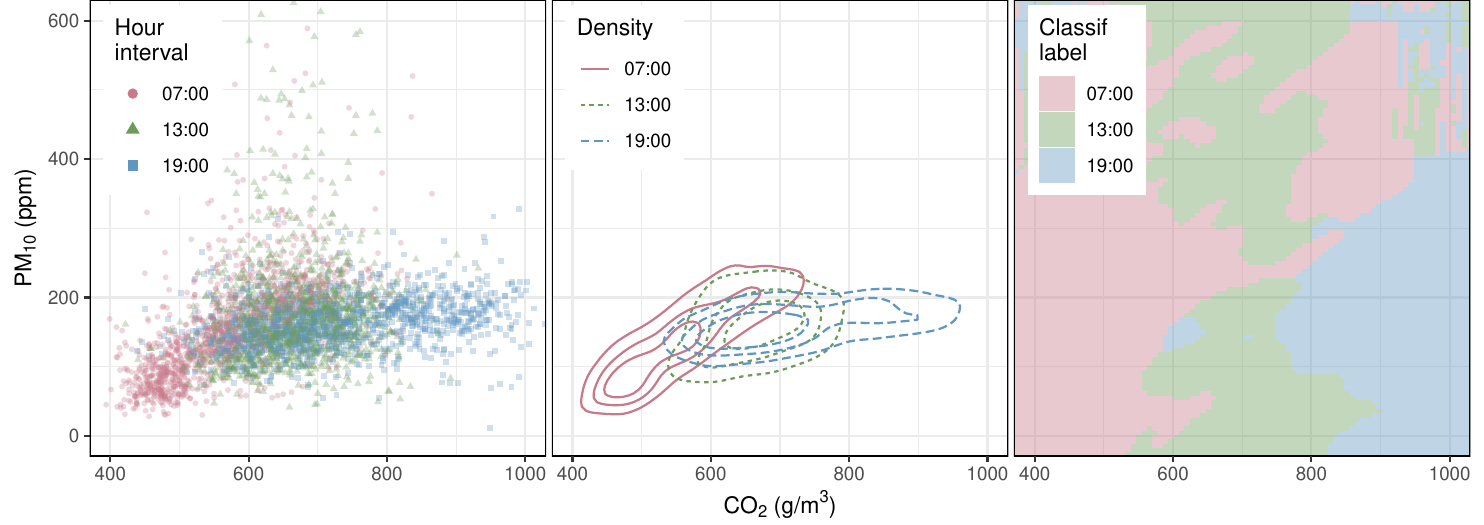} 
\caption{Density-based classifier for air quality measurements at different hourly intervals. (Left) Scatter plots for 07:00--08:00 ($n_1=1282$), 13:00--14:00 ($n_2=1280$), 19:00--20:00 ($n_3=1292$). (Centre) Quartile probability contours of marginal density estimates. (Right) Class label estimates.}
\label{fig:kda1}
\end{figure}

Our data sample comprises the  air quality measurements for three hourly intervals at six hours apart throughout the day, i.e. 07:00--08:00 (pink circles, $n_1=1282$), 13:00--14:00 (green triangles, $n_2=1280$), and 19:00--20:00 (blue squares, $n_3=1292$), as shown in the scatter plot in the left panel of Figure~\ref{fig:kda1}. In the centre panel are the quartile probability contour plots of marginal density estimates $\hat{\pi}_1 \hat{f}_1$ (pink solid lines), $\hat{\pi}_2 \hat{f}_2$ (green dotted lines), and $\hat{\pi}_3 \hat{f}_3$ (blue dashed lines), where $\hat{f}_1$ is the density estimate for 07:00--08:00, $\hat{f}_2$ for 13:00--14:00, and $\hat{f}_3$ for 19:00--20:00. As the marginal density contours have considerable overlap in the central regions, it is difficult to decide visually which marginal density value is higher. This is resolved in the plot of estimated class labels from the density-based classifier on the right of Figure~\ref{fig:kda1}. The regions where the 07:00--08:00 measurements are more likely are coloured in pink, the 13:00-14:00 measurements are more likely are in green, and the 19:00--20:00 measurements are more likely are in blue.  The general trend is, as the day progresses, both levels of CO$_2$ and PM$_{10}$ increase, with the increase of PM$_{10}$ being more sustained. The boundaries of these class label regions are complex and would not be well-estimated by the linear and quadratic classifiers. It appears that the upper right corner of the kernel classifier gives noisy decision boundaries but since this region has no observed data, it is has little effect on its accuracy. 
The misclassification rate of the kernel classifier is 0.38, in comparison to 0.44 for a linear classifier (\code{MASS::lda}) and 0.43 for a quadratic classifier (\code{MASS::qda}). 

The command to compute a tidy kernel classifier is \code{tidy\_kda}. It requires a grouped tidy tibble as its input (\code{air\_gr}), grouped by the class factor variable (\code{time}). To produce the marginal densities plot for the density-based classifier in the centre panel in Figure~\ref{fig:kda1}:  
\begin{verbatim}
R> ## tidy density-based classifier contours
R> air <- group_by(air, time)
R> air_gr <- filter(air, time %in% c("07:00", "13:00", "19:00"))
R> air_gr <- mutate(air_gr, time=droplevels(time))
R> t4 <- tidy_kda(air_gr)
R> ggplot(t4, aes(x=co2, y=pm10)) + geom_contour_ks(aes(colour=time)) 
\end{verbatim}
The layer function \code{geom\_contour\_ks} draws the contour lines for \code{tidy\_ks} objects. In addition to the columns already present in the density estimate, the extra columns in the output of a density-based classifier relate to the classes: \code{prior\_prob} (class sample proportion), \code{label} (estimated class label), \code{time} (same as input class label). The structure of a density-based classifier is similar to that for a density estimate grouped by a class variable.

\subsection{Geospatial classification}
\begin{figure}[!ht]
\centering
\includegraphics[width=0.95\textwidth]{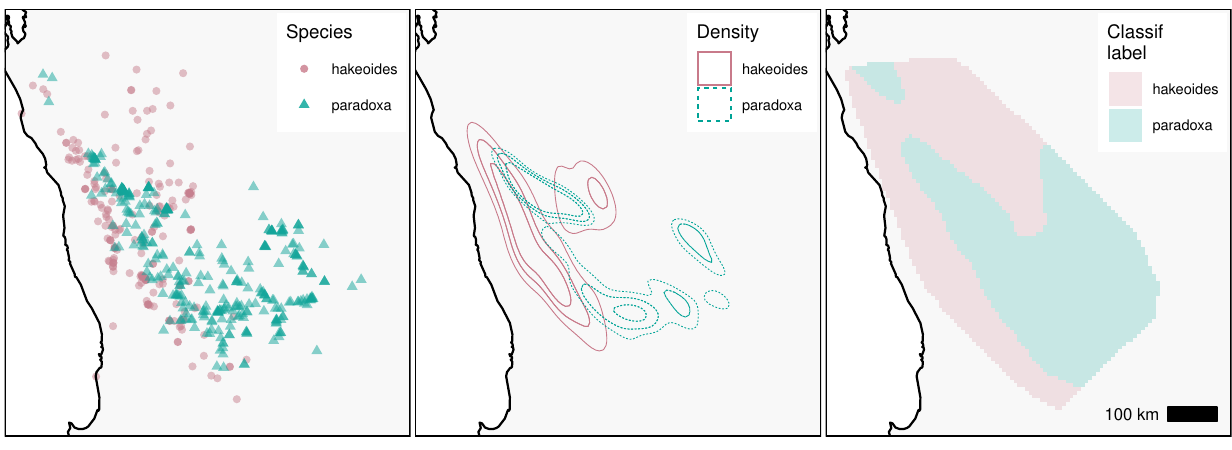} 
\caption{Density-based classifier for {\it Grevillea} locations. (Left) Scatter plots for {\it G. hakeoides} ($n_1=207$), {\it G. paradoxa} ($n_2=358$). (Centre) Quartile probability contours of marginal density estimates. (Right) Class label estimates.}
\label{fig:kda2}
\end{figure}

Our geospatial data sample comprises the combined {\it G. hakeoides} (pink circles, $n_1=207$) and {\it G. paradoxa} (green triangles, $n_2=358$) locations, as shown in the scatter plot in the left panel of Figure~\ref{fig:kda2}. In the centre panel are the quartile probability contour plots of marginal density estimates $\hat{\pi}_1 \hat{f}_1$ (pink solid lines) and $\hat{\pi}_2 \hat{f}_2$ (green dotted lines), where $\hat{f}_1$ is the density estimate for {\it G. hakeoides}, and $\hat{f}_2$ for {\it G. paradoxa}. The regions where {\it G. hakeoides} is more likely are coloured in pink, and where {\it G. paradoxa} is more likely are in green. For display purposes, the class labels have been truncated to the convex hull of the marginal density estimates so that they remain over the land area (in grey). We observe again that boundaries of these class label regions are complex and would not be well-estimated by the linear and quadratic classifiers.
 
A geospatial density-based classifier requires the input (\code{grevilleasf\_gr}) to be grouped by the class factor variable (\code{species}):
\begin{verbatim}
R> ## geospatial density-based classifier
R> grevilleasf_gr <- filter(grevilleasf, species %in% c("paradoxa","hakeoides"))
R> grevilleasf_gr <- mutate(grevilleasf_gr, species=factor(species))
R> grevilleasf_gr <- group_by(grevilleasf_gr, species)
R> s4 <- st_kda(grevilleasf_gr)
\end{verbatim}
The estimated class labels are stored in the \code{sf\_ks} object in the \code{grid} field as a collection of rectangular polygons. To plot these class labels, as in the right panel in Figure~\ref{fig:kda2}, we call \code{ggplot2::geom\_sf} on the \code{grid} field for a \pkg{ggplot2} plot, and \code{plot(x, which\_geometry=="grid")} for a base \proglang{R} plot. 
\begin{verbatim}
R> ## geospatial density-based classifier geom_sf plot 
R> ggplot(s4) + geom_sf(data=s4$grid, aes(fill=label), alpha=0.2, colour=NA) 
R> ## base R plot
R> plot(s4, which_geometry="grid", border=NA)
\end{verbatim}

The question of optimal bandwidths for a density-based classifier is more complicated than that for a density estimate. We opt for a simple and robust implementation in the \pkg{eks} package, where \code{tidy\_kda} and \code{st\_kda} call \code{ks::Hpi} for each class data sub-sample. These class-wise optimal bandwidths are known to asymptotically minimise the misclassification error.  Whilst there is an intuitive appeal in selecting bandwidths to exactly minimise the misclassification error, it is not clear how much is gained in practise with this more complicated approach over the simpler bandwidths. Moreover, there are currently no efficient algorithms to compute these more complicated  bandwidths. See \citet[Section~7.2]{chacon2018} for a discussion.

\section{Density derivative estimation} \label{sec:kdde}

Crucial information about the structure of a data set is not always revealed by examining solely the density values, and can only be discerned via the density derivatives. For example, the local minima/maxima of the data density are characterised as the locations where the first derivative is identically zero. A recent example of the utility of density derivative estimates in data analysis is the segmentation of digital images, which utilised the first density derivative of pixel colour-locations to guide the search for similar image segments more efficiently than using only the density of the pixel colour-locations \citep{beck2016}. 

With the same data as for the density estimation case, i.e., $\bX_1, \dots, \bX_n$ is a random sample drawn from the common density function $f$, our goal is to estimate the first (gradient) derivative of the unknown density $f$. For 2-dimensional data, the gradient of a density function $f$ is comprised of two partial derivatives $\D f = [\partial f / \partial x_1,  \partial f / \partial x_2]$. The kernel estimate of the density gradient is given by 
\begin{equation}
\D \hat{f}_\bH(\bx) = n^{-1} \sum_{i=1}^n \D K_\bH(\bx - \bX_i)
\label{eq:kdde}
\end{equation}
where the gradient kernel function is $\D K_\bH (\bx) = -(2\pi)^{-1} |\bH|^{-1/2} \bH^{-1} \bx \exp (-\tfrac12 \bx^\top \bH^{-1} \bx)$. 

\subsection{Tidy density derivative estimation}

\begin{figure}[!ht]
\centering
\begin{tabular}{@{}c@{}c@{}}
\includegraphics[width=0.38\textwidth,align=t]{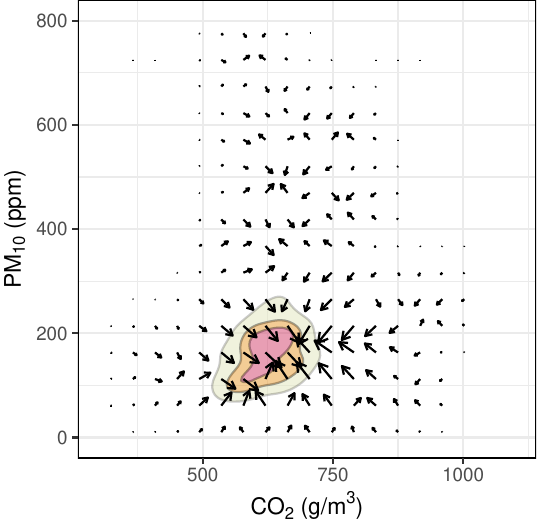} &
\includegraphics[width=0.311\textwidth,align=t]{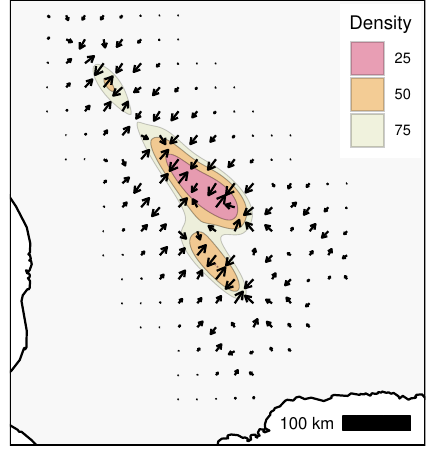} 
\end{tabular}
\caption{Quiver plots of density gradient estimate, superposed over density estimates. (Left) Air quality measurements for 13:00--14:00 ($n=1280$). (Right) {\it G. yorkrakinensis} locations ($n=93$).}
\label{fig:kdde1}
\end{figure}

Since there are two components of the density gradient, it can be visualised using two separate plots, one for each partial derivative. A more concise alternative is a quiver plot, in which arrows, whose length and direction are determined by the gradient, are drawn at each point in the estimation grid. The right panel of Figure~\ref{fig:kdde1} is the quiver plot for the density gradient estimate for the air quality measurements from 13:00 to 14:00, superposed on the density estimate. The arrows for the density gradient point towards the peaks of the modal regions. These arrows are longer where the density gradient is steeper, and they are shorter in the density tails where the slope is flatter. These density gradients indicate the rate of change in the data density, which is not easy to ascertain from the density levels themselves in the underlying density contour plot.    

The command for a tidy density gradient estimate is \code{tidy\_kdde(x, deriv\_order=1)}. The function \code{tidy\_kquiver} converts the output from \code{tidy\_kdde} into a format suitable for the quiver plot layer function \code{ggquiver::geom\_quiver} \citep{ggquiver}. The code to produce a quiver plot superposed on a density estimate is
\begin{verbatim}
R> ## tidy density gradient estimate 
R> air3 <- ungroup(filter(air, time=="13:00"))
R> air3 <- select(air3, co2, pm10)
R> t5 <- tidy_kdde(air3, deriv_order=1)
R> t6 <- tidy_kquiver(t6, thin=9)
R> ggplot(t1, aes(x=co2,y=pm10)) + ggquiver::geom_quiver(data=t6, aes(u=u,v=v))
\end{verbatim}
The output from \code{tidy\_kdde} is a tidy tibble which is grouped by \code{deriv\_group}.
The columns present in a density estimate are also present in a density derivative estimate, along with some additional columns relating to the derivative: \code{deriv\_order} (derivative order, 1 for the gradient), \code{deriv\_ind} (partial derivative enumeration, from 1 to 2), \code{deriv\_group} (partial derivative indices (1,0), (0,1) which correspond to $\partial/\partial x_1,  \partial/ \partial x_2$ respectively). 

\subsection{Geospatial density derivative estimation}

For geospatial data, the left panel in Figure~\ref{fig:kdde1} is the quiver plot for the density gradient estimate for {\it G. yorkrakinensis}, superposed on the density estimate. Whilst with \code{st\_kquiver} we can compute a geospatial output, \code{ggplot2::geom\_sf} is not able plot arrows, and it is not possible to overlay a \code{ggquiver::geom\_quiver} layer over a \code{geom\_sf} layer. The current work-around is to overlay a \code{ggplot2::geom\_segment} layer over a \code{geom\_sf} layer, with some trial and error required in \code{grid::arrow} to produce suitable arrows. 
\begin{verbatim}
R> ## geospatial density gradient estimate geom_sf plot
R> s5 <- st_kdde(yorkr, deriv_order=1) 
R> s6 <- st_kquiver(s5, thin=9) 
R> ggplot(s1) + geom_segment(data=s6$sf, aes(x=lon, xend=lon_end, y=lat, 
+  yend=lat_end), arrow=grid::arrow(length=0.1*s6$sf$len)) 
\end{verbatim}
On the other hand, for a base \proglang{R} plot, the display of geospatial and tidy data are freely interchangeable, so we can overlay the quiver plot \code{plot(x, display="quiver")} for a kernel density gradient estimate from the \pkg{ks} package.
\begin{verbatim}
R> ## geospatial density gradient estimate base R plot
R> plot(s6$tidy_ks$ks[[1]], display="quiver")
\end{verbatim}

For optimal bandwidth selection for kernel density gradient estimates, it is crucial to note that the optimal bandwidth matrix for $\D \hat{f}_\bH$ is not the same as that for 
$\hat{f}_\bH$. For a density estimate the optimality criterion is $M(\bH) = \int \E [\hat{f}_\bH (\bx) - f(\bx)]^2 \, \diff \bx$, whereas the criterion for a density gradient estimate is $M_1(\bH) = \int \E \lVert \D \hat{f}_\bH (\bx) - \D f(\bx) \lVert^2 \, \diff \bx$. Since $M \neq M_1$, their minimisers are also not equal in general.  The default optimal bandwidth for the density gradient estimate in the \pkg{eks} package is the plug-in bandwidth \citep{chacon2010} obtained from a call to \code{ks::Hpi(x, deriv.order=1)}. For the air quality measurements for 13:00 to 14:00, this bandwidth matrix is [441.0, 59.5; 59.5, 305.0]. In comparison, the optimal bandwidth matrix for the density estimate is [495.0, 88.0; 88.0, 245.0]. For the {\it G. yorkrakinensis} data, the optimal plug-in bandwidth matrix for the density gradient estimate is [4.39e8, $-4.36$e8; $-4.36$e8, 7.73e8], whereas for the density estimate, it is [8.84e8, $-8.33$e8; $-8.33$e8, 1.36e9]. 
 
\section{Density-based clustering (unsupervised learning)} \label{sec:kms}
The goal of clustering is to discover homogeneous groups within a data set in a trade-off between similarity/dissimilarity: members of the same cluster are similar to each other while members of different clusters are dissimilar to each other. If the $q$ unknown population clusters are $\{C_1,\dots,C_q\}$, then the cluster labelling function is $c(\bx)=j$ whenever a candidate point $\bx$ belongs to cluster $C_j$. Whilst we are able to estimate the cluster labelling function for all candidate points, for the vast majority of data analysis cases, it is sufficient to compute $\hat{c}(\bX_1), \dots, \hat{c}(\bX_n)$ for the data sample $\bX_1, \dots, \bX_n$. Since the cluster labels are unknown, then this is an unsupervised learning problem. 

Many clustering algorithms have been proposed in the literature. Our chosen approach is density-based clustering, where a cluster is a data-rich region (high density values) which is separated from another data-rich region by a data-poor region (low density values). Thus we associate each data point to its `most representative' data-rich region. In the \pkg{eks} package, this association is carried out with a mean shift algorithm \citep{fukunaga1975}. For a data point $\bX_i$, we initialise a sequence with $\bX_{i,0} = \bX_i$, then we iterate the recurrence equation
$$
\bX_{i, k+1} = \bX_{i,k} + \bH^{-1} \D \hat{f}_\bH (\bX_k)\big/\hat{f}_\bH(\bX_k),
$$where $\hat f_\bH$ is a density estimate and $\D \hat{f}_\bH$ is a density gradient estimate. This recurrence equation is closely related to the well-known gradient ascent algorithm, with the improvement that accelerates the convergence of the recurrence iterations in regions of low data density. A more computationally stable form of the  mean shift recurrence equation, since it avoids the explicit computation of the density estimate $\hat{f}_\bH$ and density derivative estimate $\D \hat{f}_\bH$, is
\begin{equation} \label{eq:kms}
\bX_{i,k+1} = \bX_{i,k} + \bbeta_\bH(\bX_{i,k})
= \frac{\sum_{\ell=1}^n\bX_\ell \, g\big((\bX_{i,k} -\bX_\ell)^\top\bH^{-1}(\bX_{i,k}-\bX_\ell)\big)}{\sum_{\ell=1}^n g \big((\bX_{i,k} - \bX_\ell)^\top\bH^{-1}(\bX_{i,k}-\bX_\ell)\big)}
\end{equation}
where $g(x) = x \exp(-\tfrac12 x)$ and $\bbeta_\bH(\bx) = \frac{\textstyle \sum_{\ell=1}^n \bX_\ell g((\bx - \bX_\ell)^\top\bH^{-1}(\bx - \bX_\ell))}{\textstyle \sum_{\ell=1}^n g((\bx - \bX_\ell)^\top\bH^{-1}(\bx - \bX_\ell))} - \bx$. This $\bbeta_\bH$ is known as the mean shift, since it is the difference between the current iterate and a weighted mean of all data points.  
For our stopping rule, we iterate the recurrence in Equation~\eqref{eq:kms} until either we reach a maximum number of iterations (400) or that the distance between subsequent iterations is less than 0.001 times the minimal marginal IQR (interquartile range) of the input data. This heuristic stopping rule gives sensible results in most cases. 

The result is a sequence of points $\{\bX_{i,0}, \bX_{i,1}, \dots\}$ which traces out a path, along the steepest ascent of the density gradient, from the data point $\bX_i$ to the mode of the associated data-rich region. The data-rich regions are the `basins of attraction' of the density gradient ascent. If the data points are associated with the same mode, then they are considered to be members of the same cluster. Thus the number of clusters is equal to the number of these basins of attraction. For more details on mean shift and other forms of density-based clustering, see \citet[Section~6.2]{chacon2018}.

\subsection{Tidy clustering}

\begin{figure}[!ht]
\centering
\includegraphics[width=0.7\textwidth]{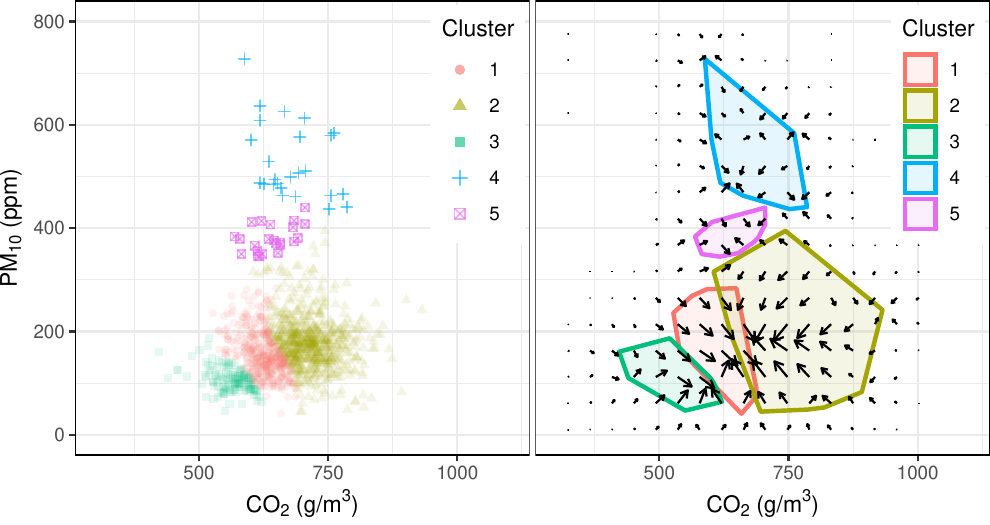} 
\caption{Mean shift clusters for the air quality measurements 13:00--14:00 ($n=1280$). (Left) Cluster members. (Right) Cluster convex hulls, superposed over the quiver plot of its density gradient estimate.}
\label{fig:kms1}
\end{figure}

The result of the mean shift clustering on the $n=1280$ air quality measurements from 13:00 to 14:00 into 5 clusters is displayed on the left panel in Figure~\ref{fig:kms1}.  Observe that we do not need to specify the number of clusters in advance, and the clusters can be of any arbitrary shape. These represent two important advantages over $k$-means clustering, which requires an a priori number of clusters, and whose cluster shapes are more restricted than those in mean shift clustering. Cluster \#4 (blue crosses) and \#5 (magenta boxed crosses) are the most separate from the other clusters. The points at the edges of cluster \#3 (green squares), cluster \#1 (red circles)  and cluster \#2 (khaki triangles) are close to together, and $k$-means clustering tends to assign them to the same cluster, whereas the directionality of the mean shift assigns them to different clusters. Since the mean shift relies on the density gradient ascent paths, we overlay the arrows of the quiver plot of the density gradient on the convex hulls of the mean shift clusters on the right of Figure~\ref{fig:kms1}. We observe that the gradient ascent arrows within each cluster are oriented towards the modes.

The command for mean shift clustering for tidy data is \code{tidy\_kms}. The output is similar to that for a single density estimate, except that the data points are returned rather than the estimation grid points, and that \code{estimate} indicates the estimated cluster label rather than the density estimate value. 
\begin{verbatim}
R> ## tidy mean shift clusters 
R> t7 <- tidy_kms(air3)
R> ggplot(t7, aes(x=co2, y=pm10)) + geom_point(aes(colour=estimate)) 
\end{verbatim}

Since the direction along which the data points are shifted is directly related to the density gradient, the default bandwidth for mean shift clustering in  \code{tidy\_kms} is the plug-in bandwidth computed by \code{ks::Hpi(x, deriv.order=1)}. For the air quality measurements for 13:00 to 14:00, this bandwidth matrix is [441.0, 59.5; 59.5, 305.0]. The bandwidth choice is made with the goal of optimal identification of the density gradient ascent paths. It is also supported by the results that the optimal bandwidth for estimating the mode of a density is closely related to the optimal bandwidth for density gradient estimation \citep[p.~138]{chacon2018}. 

\subsection{Geospatial clustering}

\begin{figure}[!ht]
\centering
\includegraphics[width=0.66\textwidth]{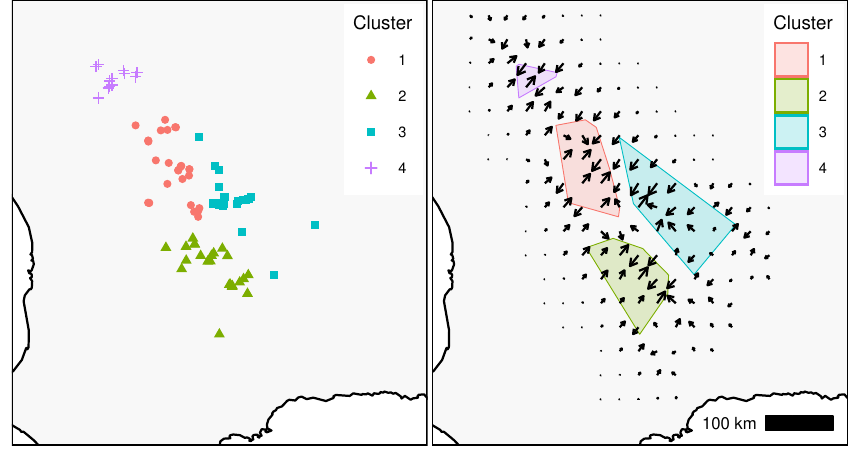} 
\caption{Mean shift clusters for {\it G. yorkrakinensis} ($n=93$). (Left) Cluster members. (Right) Cluster convex hulls, superposed over the quiver plot of its density gradient estimate.}
\label{fig:kms}
\end{figure}

The result of the mean shift clustering on the $n=93$ {\it G. yorkrakinensis} locations into 4 clusters is displayed on the left panel in Figure~\ref{fig:kms}. Cluster \#4 (magenta crosses) is the most northerly and most separate from the other clusters. Cluster \#2 (green triangles) forms the most southerly cluster and is also well-separated. The points on the right edge of cluster \#1 (red circles) are close to those on the left edge of cluster \#3 (cyan squares), though the directionality of the mean shift, as indicated by the black arrows of the density gradient, assigns them to different clusters.  

The command for mean shift clustering for geospatial data is \code{st\_kms}. To produce the mean shift clusters, with at least 3 members in each cluster, the code snippet is: 
\begin{verbatim} 
R> ## geospatial mean shift clusters geom_sf plot 
R> s7 <- st_kms(yorkr, min.clust.size=3)
R> ggplot(s7) + geom_sf(aes(colour=estimate)) 
R> ## base R plot
R> plot(s7, pch=16)
\end{verbatim}
The default bandwidth for mean shift clustering in \code{st\_kms} is the plug-in bandwidth computed by \code{ks::Hpi(x, deriv.order=1)}. For the {\it G. yorkrakinensis} locations, this is [4.39e8, $-$4.36e8; $-$4.36e8, 7.73e8]. 

\section{Software} \label{sec:other}

\subsection{Other data analysis settings}

All the functionality in the \pkg{ks} package that involve 1- and 2-dimensional kernel smoothers are 
implemented for tidy data in the \pkg{eks} package. In addition to \code{tidy\_kde}, \code{tidy\_kda}, \code{tidy\_kdde}, and \code{tidy\_kms} in Sections~\ref{sec:kde}--\ref{sec:kms}, these functions include 
\begin{description} 
\item[\code{tidy\_kde\_boundary}] Boundary density estimate where the kernel function $K$ is modified explicitly in the boundary region 
\item[\code{tidy\_kde\_truncate}] Truncated density estimate where the standard density estimate $\hat{f}$ is truncated and rescaled to give unit integral over the boundary region 
\item[\code{tidy\_kde\_sp}] Sample point density estimate where the bandwidth $\bH(\cdot)$ varies with the data point $\bX_i$
\item[\code{tidy\_kde\_balloon}] Balloon density estimate where the bandwidth $\bH(\cdot)$ varies with the estimation point $\bx$
\item[\code{tidy\_kdcde}] Deconvolved density estimate for data $\bX_i$ observed with error 
\item[\code{tidy\_kcde}] Cumulative distribution estimate $\hat{F}$
\item[\code{tidy\_kcopula}] Copula estimate with uniformly distributed marginal distributions
\item[\code{tidy\_kroc}] ROC (receiver operating characteristic) curve of the 2-sample comparison of the  marginal distribution estimates $\hat{F}_1, \hat{F}_2$
\item[\code{tidy\_kdr}] Density ridge estimate which is a generalisation of principal components to lower dimensional manifolds  
\item[\code{tidy\_kde\_local\_test}] Significance testing for the 2-sample comparison of the difference of the density estimates $\hat{f}_1 - \hat{f}_2$
\item[\code{tidy\_kfs}] Significance testing for modal regions where the second derivative (Hessian matrix) of the density estimate $\hat{f}$ is positive definitive. 	
\end{description}  
All of the above functions (except \code{tidy\_kcopula}) are implemented for 2-dimensional geospatial data as \code{st\_k*}. All of these utilise the appropriate default bandwidth selector from the \pkg{ks} package. For brevity, we do not illustrate them here: their usage is demonstrated in their help pages contained in the \pkg{eks} package, and the details of the statistical framework in these data analysis settings are provided in \citet{chacon2018}.

\subsection{Export to external GIS}
The ability to export the geospatial kernel estimates to standard vectorial geospatial data formats extends the functionality of the \pkg{eks} package to GIS software. The commands to export to the geopackage format are:   
\begin{verbatim}
R> ## export to vectorial geospatial format
R> sf::write_sf(yorkr, dsn="grevillea.gpkg", layer="yorkr")
R> sf::write_sf(st_get_contour(s1), dsn="grevillea.gpkg", layer="yorkr_cont")
R> sf::write_sf(s6$sf, dsn="grevillea.gpkg", layer="yorkr_quiver")
\end{verbatim}
The \code{grevillea.gpkg} geopackage consists of four layers: \code{yorkr} for the point geometries of the {\it G. yorkrakinensis} locations, \code{yorkr\_cont} for the multi-polygons of the quartile contour regions of the density estimate, and \code{yorkr\_quiver} for the linestrings of the density gradient flows. 

\begin{figure}[!ht]
\centering
\includegraphics[width=0.4\textwidth]{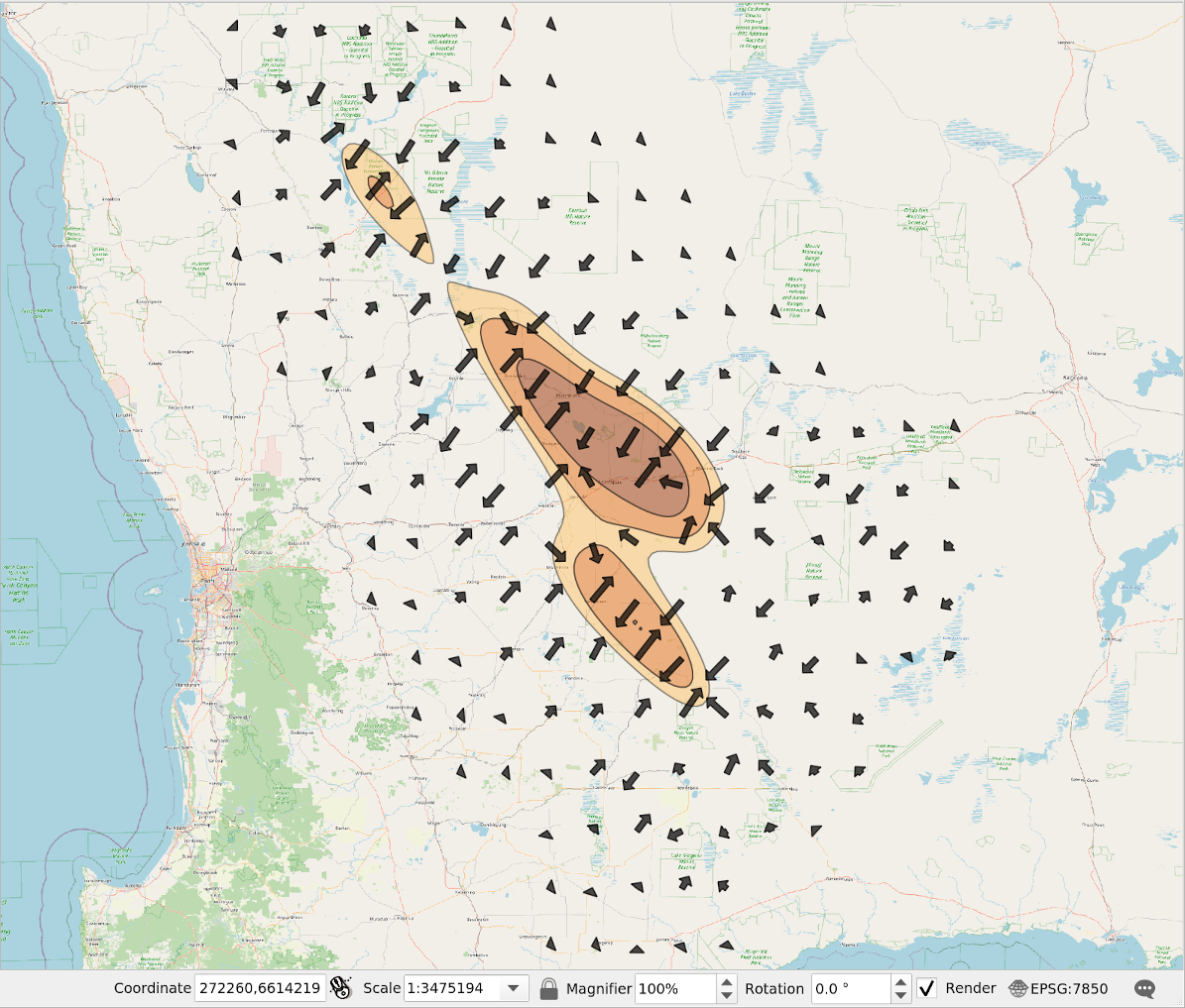} 
\includegraphics[width=0.4\textwidth]{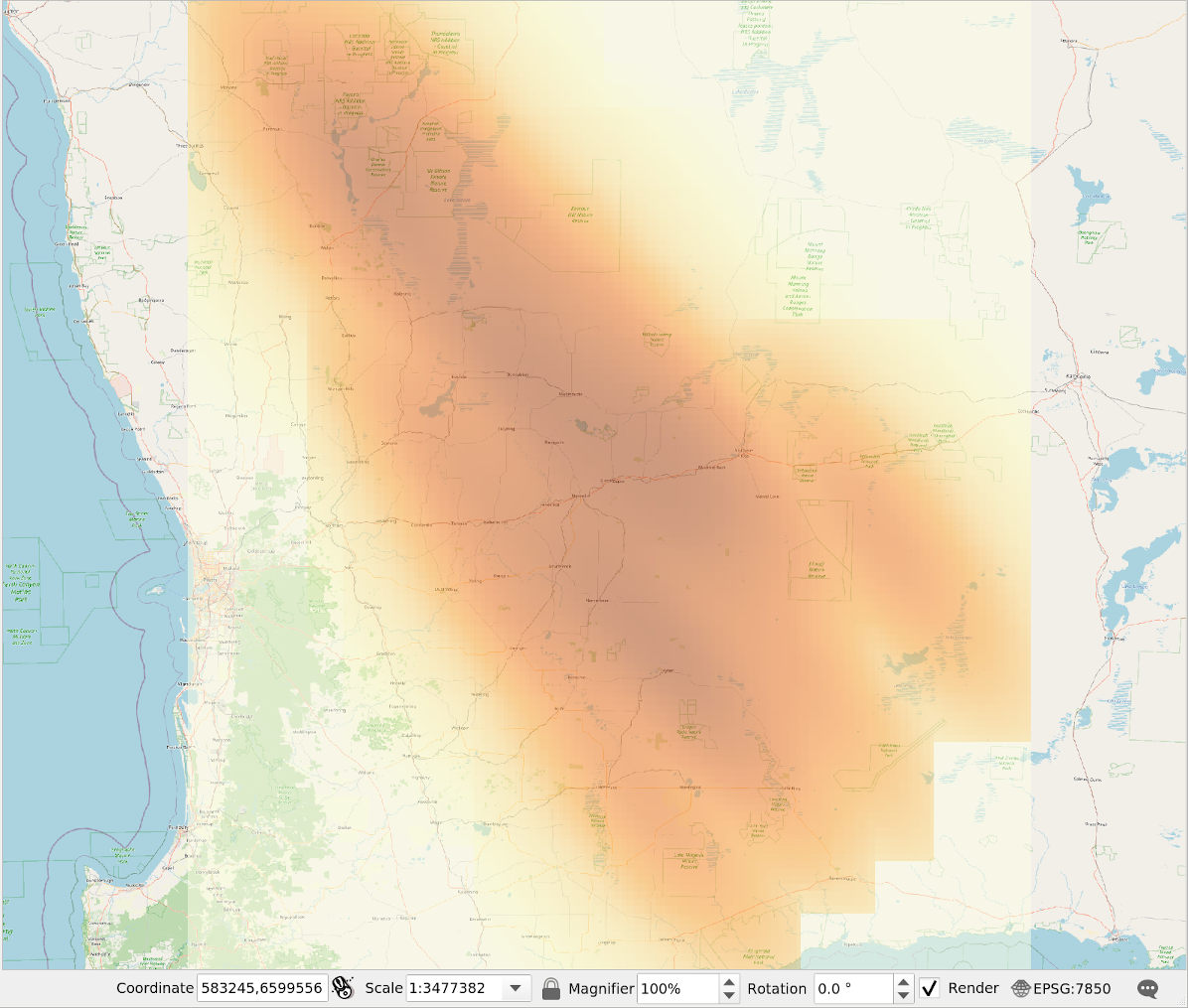} 
\caption{Screenshots of QGIS analysis for {\it G. yorkrakinensis} ($n=93$). (Left) Contour plot of  vectorial density estimate and quiver plot of vectorial density gradient estimate. (Right) Heat map of raster density estimate.}
\label{fig:qgis}
\end{figure}

This \code{grevillea.gpkg} geopackage can be subsequently employed in QGIS \citep{qgis}, which is an industry standard software for GIS practitioners since it offers features that are not available in \proglang{R}. For example, it has an interactive point-and-click interface, and it incorporates fast rendering of the OpenStreetMap base maps. A screenshot from a QGIS analysis for a quiver plot overlaid on a density estimate is given in the left panel of Figure~\ref{fig:qgis}. Recall that quiver plots can be difficult to produce with geospatial data in \pkg{ggplot2} graphics, since the arrows require trial and error to display suitably with \code{ggplot2::geom\_segment}. In contrast, quiver plots are straightforward in QGIS since rescaleable arrows are a native feature.

In addition, QGIS efficiently handles raster geospatial data. Whilst the \code{grid} field of a kernel estimate consists of the rectangular polygons for each pixel of the estimation grid, it can be converted to a raster via the \pkg{stars} package. The heat map of the converted raster is displayed in QGIS on the right of Figure~\ref{fig:qgis}.  
\begin{verbatim}
R> ## export to raster format
R> stars::write_stars(stars::st_rasterize(s1$grid), dsn="grevillea.gpkg", 
   options=c("APPEND_SUBDATASET=YES", "RASTER_TABLE=yorkr_raster"))
\end{verbatim}
 
\section{Conclusion}

We have introduced a new \proglang{R} package \pkg{eks} which serves as a bridge from the comprehensive suite of kernel smoothers in the \pkg{ks} package to the tidyverse and geospatial analysis. A wide range of kernel smoothing methods are available, which (i) improve on the existing kernel density estimates, and (ii) widen the accessibility to more complex kernel-based data analyses, such as density gradient estimation, density-based classification (supervised learning) and mean shift clustering (unsupervised learning). The \pkg{eks} package provides practitioners with additional tools to create compelling statistical visualisations from kernel smoothers, whether they are using tidy or geospatial data, or whether they are using base \proglang{R} or tidyverse graphics. 


\begin{thebibliography}{}
\bibitem[\protect\citeauthoryear{Ba\'illo and Chac\'on}{Ba\'illo and
	Chac\'on}{2021}]{baillo2021}
Ba\'illo, A. and J.~E. Chac\'on (2021).
\newblock Chapter 1 - statistical outline of animal home ranges: An application
of set estimation.
\newblock In A.~{Srinivasa Rao} and C.~Rao (Eds.), {\em Data Science: Theory
	and Applications}, Volume~44 of {\em Handbook of Statistics}, pp.\  3--37.
Elsevier.

\bibitem[\protect\citeauthoryear{Beck, Duong, Azzag, and Lebbah}{Beck
	et~al.}{2016}]{beck2016}
Beck, G., T.~Duong, H.~Azzag, and M.~Lebbah (2016).
\newblock Distributed mean shift clustering with approximate nearest
neighbours.
\newblock In {\em Proceedings of the 2016 International Conference on Neural
	Networks (IJCNN)}, pp.\  3110--3115.

\bibitem[\protect\citeauthoryear{B\'eranger, Duong, Perkins-Kirkpatrick, and
	Sisson}{B\'eranger et~al.}{2019}]{beranger2019}
B\'eranger, B., T.~Duong, S.~E. Perkins-Kirkpatrick, and S.~A. Sisson (2019).
\newblock Tail density estimation for exploratory data analysis using kernel
methods.
\newblock {\em Journal of Nonparametric Statistics\/}~{\em 31}, 144--174.

\bibitem[\protect\citeauthoryear{Bowman and Foster}{Bowman and
	Foster}{1993}]{bowman1993}
Bowman, A.~W. and P.~Foster (1993).
\newblock Density based exploration of bivariate data.
\newblock {\em Statistics and Computing\/}~{\em 3}, 171--177.

\bibitem[\protect\citeauthoryear{Chac\'on and Duong}{Chac\'on and
	Duong}{2010}]{chacon2010}
Chac\'on, J.~E. and T.~Duong (2010).
\newblock Multivariate plug-in bandwidth selection with unconstrained bandwidth
matrices.
\newblock {\em Test\/}~{\em 19}, 375--398.

\bibitem[\protect\citeauthoryear{Chac\'on and Duong}{Chac\'on and
	Duong}{2018}]{chacon2018}
Chac\'on, J.~E. and T.~Duong (2018).
\newblock {\em Multivariate Kernel Smoothing and Its Applications}.
\newblock Boca Raton: Chapman and Hall/CRC.

\bibitem[\protect\citeauthoryear{Chac\'on, Duong, and Wand}{Chac\'on
	et~al.}{2011}]{chacon2011ss}
Chac\'on, J.~E., T.~Duong, and M.~P. Wand (2011).
\newblock Asymptotics for general multivariate kernel density derivative
estimators.
\newblock {\em Statistica Sinica\/}~{\em 21}, 807--840.

\bibitem[\protect\citeauthoryear{Devroye, Gy\"orfi, and Lugosi}{Devroye
	et~al.}{1996}]{devroye1996}
Devroye, L., L.~Gy\"orfi, and G.~Lugosi (1996).
\newblock {\em A Probabilistic Theory of Pattern Recognition}.
\newblock Springer.

\bibitem[\protect\citeauthoryear{Duong}{Duong}{2007}]{duong2007}
Duong, T. (2007).
\newblock {ks}: {K}ernel density estimation and kernel discriminant analysis
for multivariate data in {R}.
\newblock {\em Journal of Statistical Software\/}~{\em 21(7)}, 1--16.

\bibitem[\protect\citeauthoryear{Duong}{Duong}{2023}]{eks}
Duong, T. (2023).
\newblock {\em {eks}: {T}idy and Geospatial Kernel Smoothing}.
\newblock R package version 1.0.3.

\bibitem[\protect\citeauthoryear{Duong and Hazelton}{Duong and
	Hazelton}{2003}]{duong2003}
Duong, T. and M.~L. Hazelton (2003).
\newblock Plug-in bandwidth matrices for bivariate kernel density estimation.
\newblock {\em Journal of Nonparametric Statistics\/}~{\em 15}, 17--30.

\bibitem[\protect\citeauthoryear{Fukunaga and Hostetler}{Fukunaga and
	Hostetler}{1975}]{fukunaga1975}
Fukunaga, K. and L.~Hostetler (1975).
\newblock The estimation of the gradient of a density function, with
applications in pattern recognition.
\newblock {\em {IEEE} Transactions on Information Theory\/}~{\em 21}, 32--40.

\bibitem[\protect\citeauthoryear{Hyndman}{Hyndman}{1996}]{hyndman1996jasa}
Hyndman, R.~J. (1996).
\newblock Computing and graphing highest density regions.
\newblock {\em American Statistician\/}~{\em 50}, 120--126.

\bibitem[\protect\citeauthoryear{Kalair and Connaughton}{Kalair and
	Connaughton}{2021}]{kalair2021}
Kalair, K. and C.~Connaughton (2021).
\newblock Anomaly detection and classification in traffic flow data from
fluctuations in the flow–density relationship.
\newblock {\em Transportation Research Part C: Emerging Technologies\/}~{\em
	127}, 103178.

\bibitem[\protect\citeauthoryear{Myers, Mittermeier, Mittermeier, Da~Fonseca,
	and Kent}{Myers et~al.}{2000}]{myers2000}
Myers, N., R.~A. Mittermeier, C.~G. Mittermeier, G.~A.~B. Da~Fonseca, and
J.~Kent (2000).
\newblock Biodiversity hotspots for conservation priorities.
\newblock {\em Nature\/}~{\em 403}, 853--858.

\bibitem[\protect\citeauthoryear{{OGC}}{{OGC}}{2010}]{ogc-sfa}
{OGC} (2010).
\newblock {OpenGIS} implementation standard for geographic information -
{S}imple feature access - {P}art 1: Common architecture. {V}ersion 1.2.1.

\bibitem[\protect\citeauthoryear{O'Hara-Wild}{O'Hara-Wild}{2019}]{ggquiver}
O'Hara-Wild, M. (2019).
\newblock {\em {ggquiver}: {Q}uiver Plots for `ggplot2'}.
\newblock R package version 0.2.0.

\bibitem[\protect\citeauthoryear{Otto and Kahle}{Otto and
	Kahle}{2023}]{ggdensity}
Otto, J. and D.~Kahle (2023).
\newblock {\em {ggdensity}: {I}nterpretable Bivariate Density Visualization
	with `ggplot2'}.
\newblock R package version 1.0.0.

\bibitem[\protect\citeauthoryear{Pebesma}{Pebesma}{2018}]{sf}
Pebesma, E. (2018).
\newblock Simple features for {R}: Standardized support for spatial vector
data.
\newblock {\em The R Journal\/}~{\em 10}, 439--446.

\bibitem[\protect\citeauthoryear{{QGIS.org}}{{QGIS.org}}{2021}]{qgis}
{QGIS.org} (2021).
\newblock {\em {QGIS} Geographic Information System}.
\newblock QGIS Association.

\bibitem[\protect\citeauthoryear{RATP}{RATP}{2016}]{ratp2016}
RATP (2016).
\newblock Qualit\'e de l'air mesur\'ee dans la station {C}h\^atelet.
\newblock
\url{https://data.iledefrance.fr/explore/dataset/qualite-de-lair-mesuree-dans-la-station-chatelet}.
R\'egie autonome des transports parisiens -- D\'epartement D\'eveloppement,
Innovation et Territoires.
\newblock Accessed 2017-09-27.

\bibitem[\protect\citeauthoryear{Rudis, Bolker, and Schulz}{Rudis
	et~al.}{2017}]{ggalt}
Rudis, B., B.~Bolker, and J.~Schulz (2017).
\newblock {\em {ggalt}: {E}xtra Coordinate Systems, `Geoms', Statistical
	Transformations, Scales and Fonts for `ggplot2'}.
\newblock R package version 0.4.0.

\bibitem[\protect\citeauthoryear{Schauer, Duong, Bleakley, Bardin, Bornens, and
	Goud}{Schauer et~al.}{2010}]{schauer2010}
Schauer, K., T.~Duong, K.~Bleakley, S.~Bardin, M.~Bornens, and B.~Goud (2010).
\newblock Probabilistic density maps to study global endomembrane organization.
\newblock {\em Nature Methods\/}~{\em 7}, 560--568.

\bibitem[\protect\citeauthoryear{Wickham}{Wickham}{2014}]{wickham2014}
Wickham, H. (2014).
\newblock Tidy data.
\newblock {\em Journal of Statistical Software\/}~{\em 59(10)}, 1--23.

\bibitem[\protect\citeauthoryear{Wickham}{Wickham}{2016}]{ggplot2}
Wickham, H. (2016).
\newblock {\em {ggplot2}: {E}legant Graphics for Data Analysis}.
\newblock New York: Springer-Verlag.
\end{thebibliography}

\end{document}